\documentclass[twocolumn,secnumarabic,amssymb,amsmath,nofootinbib,tightenlines,aps,prb]{revtex4}
\usepackage{amsmath}
\usepackage{longtable}
\usepackage{bm}
\usepackage{epsfig}
\usepackage{mathrsfs}
\usepackage{amssymb}
\usepackage{color}
\usepackage{latexsym}
\usepackage{subfigure}
\usepackage{array}
\usepackage{gensymb}
\usepackage{textcomp}
\usepackage{hyperref}

\newcommand{\fig}[1]{Fig.~\ref{#1}}

\begin{document}

\author{Somnath~Bhowmick}
\email{bsomnath@mrc.iisc.ernet.in}
\affiliation{Materials Research Center, Indian Institute of Science, Bangalore, India}
\author{Umesh V Waghmare}
\email{waghmare@jncasr.ac.in}
\affiliation{Jawaharlal Nehru Centre for Advanced Scientific Research (JNCASR), Bangalore, India}
\title{Anisotropy of the Stone-Wales Defect and Warping of Graphene Nano-ribbons: A First-principles Analysis}
\date{\today}

\begin{abstract}
Stone-Wales (SW) defects, analogous to dislocations in crystals, play an important role in
mechanical behavior of $sp^2$-bonded carbon based materials. Here, we show using first-principles
calculations that a marked anisotropy in the interaction among the SW defects has interesting 
consequences when such defects are present near the edges of a graphene nano-ribbon: depending
on their orientation with respect to edge, they result in compressive or tensile stress,
and the former is responsible to depression or warping of the graphene nano-ribbon. Such warping
results in delocalization of electrons in the defect states.
\end{abstract}
\maketitle
\section{Introduction}
Graphene, a two dimensional (2D) allotrope of carbon, is the motherform of all its graphitic
forms. Graphene shows several fascinating electronic-transport properties, originating
from the linear energy dispersion near the high symmetry corners of the hexagonal Brillouin zone, 
which results in effective dynamics of electrons similar to massless Dirac Fermions.\cite{neto109} 
But graphene is a zero bandgap semiconductor,\cite{wallace1947} which limits its applications 
in electronic devices. In the bulk-form, a band gap can be opened up and tuned by doping graphene with boron or nitrogen~\cite{CNR} or by introducing uniaxial strain.\cite{hua2008,zhen2008}
In graphene nano-ribbons (GNRs), this can be accomplished using the
geometry of the edges: while a GNR with a zigzag edge (ZGNR) has a vanishing gap
(which opens up due to magnetic ordering), a GNR with an arm-chair edge (AGNR)~\cite{son2006}
has a nonzero gap. GNRs can be very useful for practical purposes because their bandgap
can be tuned by changing the ribbon-width.\cite{son2006} Magnetic ground state of pristine ZGNR
can be exploited to explore graphene based spintronics.\cite{young2006}

For any technological applications of graphene, understanding of its 
structural stability and mechanical behavior is crucial.
For example, deviation from the perfectly planar structure
in the form of ripples or wrinkles observed in graphene,\cite{meyer2006,fasolino2007}
can have interesting effects on electronic properties. GNRs are known to be susceptible
to structural instabilities at the edges and reconstructions.
\cite{shenoy2008,huang2009,bets2009,koskinen2008,wassmann2008,koskinen2009,caglar2009,gass2008}
Topological defects in the honeycomb carbon lattice, such as Stone-Wales (SW) defects 
(pairs of pentagons and heptagons created by 90\textdegree~rotation of a C-C 
bond~\cite{stone1986}) occur in graphene~\cite{meyer2008}
and are relevant to its structural and mechanical behavior.\cite{ana2008,Yakobson}
It is important to understand how atomic and electronic structure of quasi 1-D GNRs 
is influenced by such defects. In this work, we focus on the effects of the SW defects on
structural stability, electronic and magnetic properties of GNRs. 

Deprived of a neighbor, an atom at the edge of GNR has a dangling bond resulting in 
an edge compressive stress, which can be relieved by warping,
as analyzed by~\citeauthor{shenoy2008}~\cite{shenoy2008} using a classical potential and
interpreted with a continuum model. \citeauthor{huang2009},\cite{huang2009} on the other hand,
found using first-principles quantum mechanical simulations that such graphene with dangling bonds at the edges would rather undergo SW mediated edge reconstructions to relieve stresses, and consequently have a flat structure. Alternatively, edge stresses in GNRs can be relieved if the dangling bonds are saturated with hydrogen (H-GNR), stabilizing the planar structure relative to the warped one.\cite{huang2009} How SW defect would influence the structure of a H-GNR is not clear and we uncover this in the present work. Although SW defects cost energy,\cite{lusk2008}
they do occur in graphene~\cite{meyer2008} and are shown here to \textit{induce warping instability}
even in H-GNRs.

We organize the paper in the following manner. First, we briefly describe computational details in section~\ref{method}. A discussion follows on various stresses associated with a SW defect in bulk graphene in section~\ref{bg}. We correlate the results obtained in this section to the mechanical properties of edge reconstructed (by SW defects) GNRs. Next, in section~\ref{gnr}, we investigate the properties of such GNRs: first the issue of structural stability in section~\ref{mp}, followed by their electronic properties in section~\ref{ep}. We conclude the paper in section~\ref{con}.

\section{Method}
\label{method}
\begin{figure}
\subfigure[]{\epsfxsize=3.5truecm \epsfbox{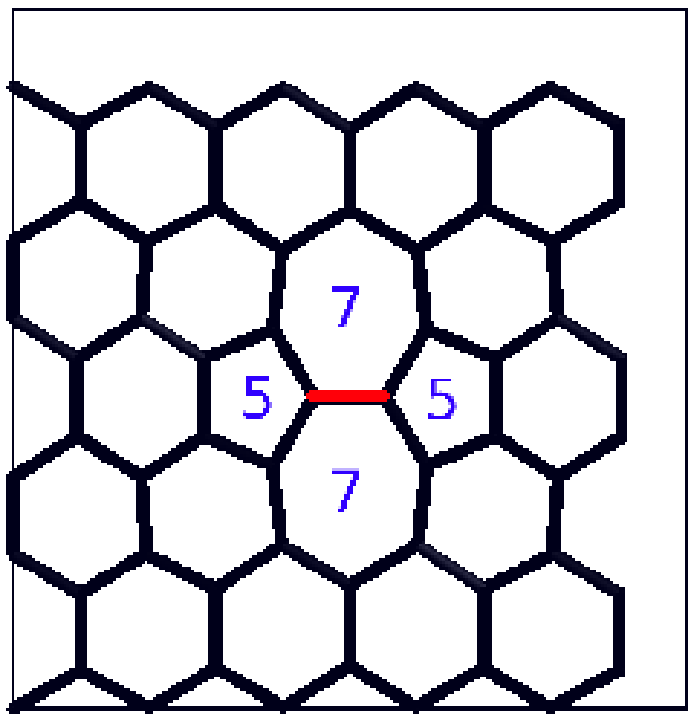}}
\subfigure[]{\epsfxsize=3.5truecm \epsfbox{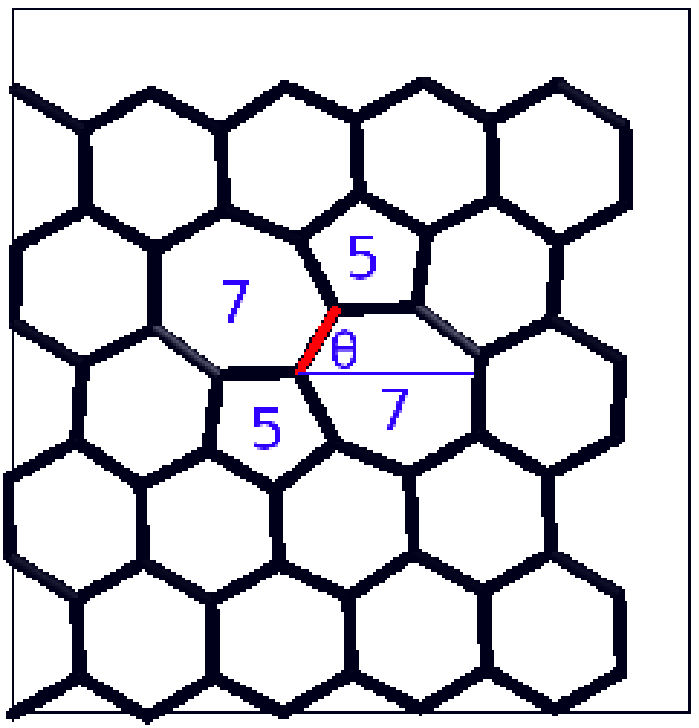}}
\subfigure[]{\epsfxsize=3.5truecm \epsfbox{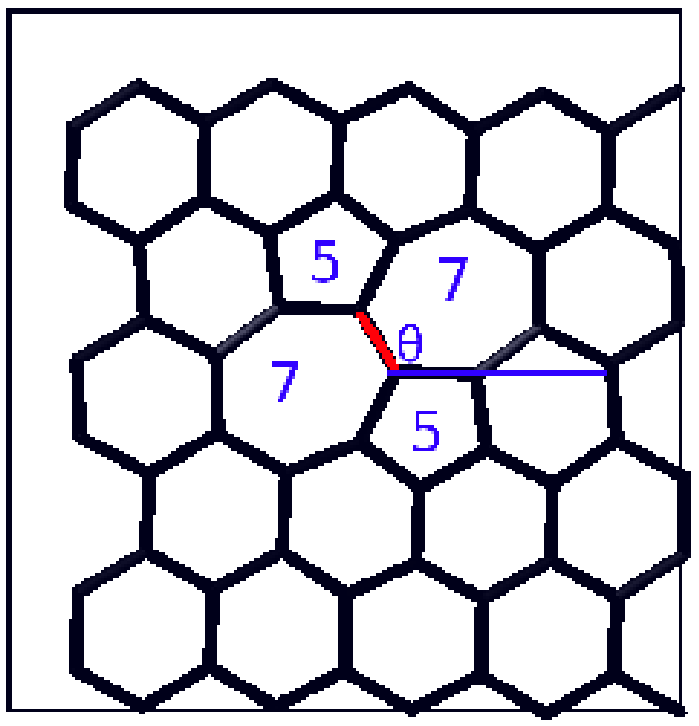}}
\subfigure[]{\epsfxsize=3.5truecm \epsfbox{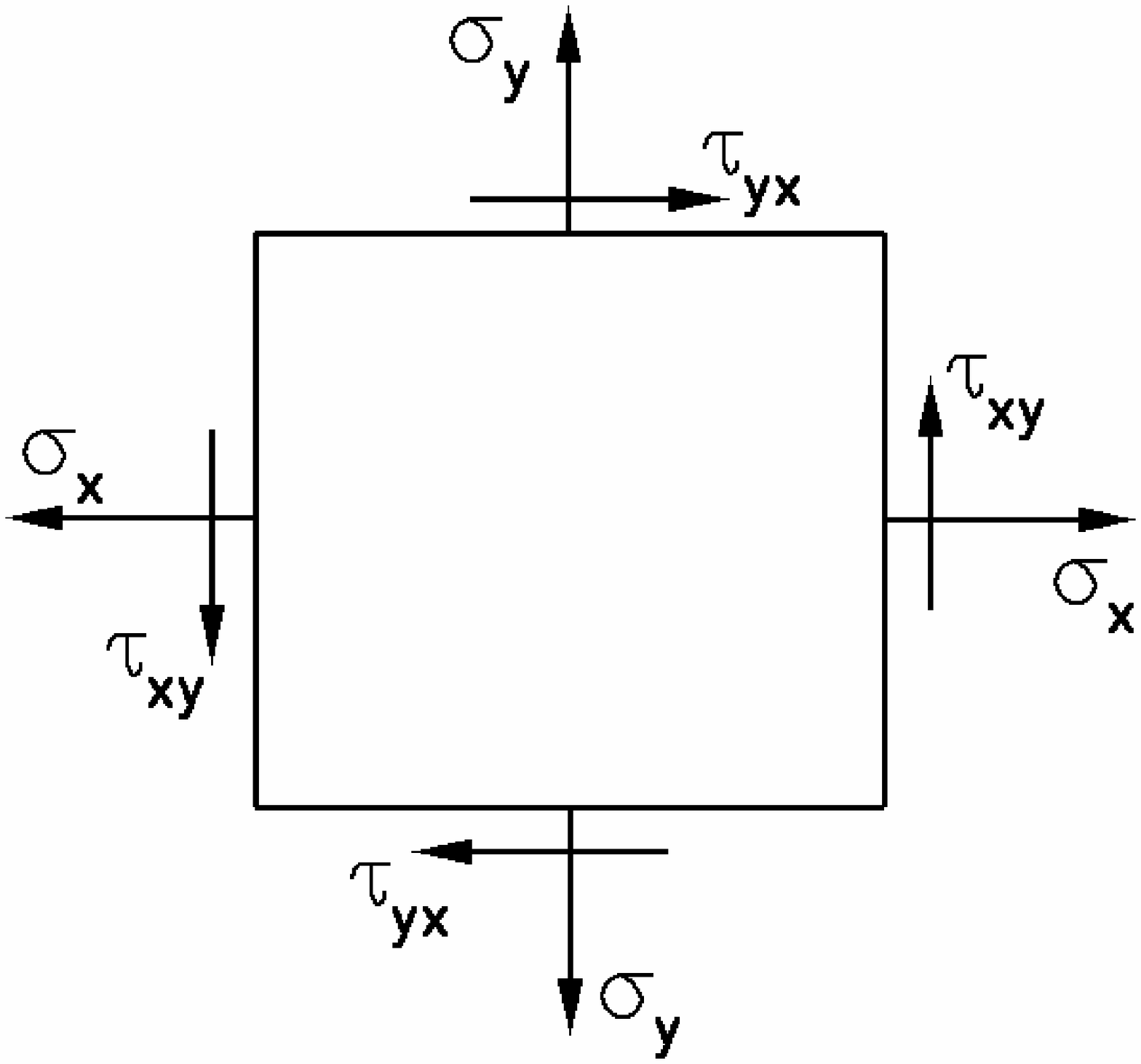}}
\caption{(color online) Stone Wales defect in bulk graphene; (a) $SW_\|$, (b) $SW_\angle$
($\theta$=60\textdegree), (c) $SW_\angle$ ($\theta$=120\textdegree), where $\theta$ is
the angle of the bond [marked in red (light gray in the gray-scale)] with the horizontal axis in the anti-clockwise direction. (d) Planar stresses acting on the supercell.}
\label{fig1}
\end{figure}
We use first-principles calculations as implemented in the PWSCF code,\cite{pwscf} with
a plane-wave basis set and ultrasoft pseudo-potential, and electron 
exchange-correlation is treated with a local density approximation (LDA, Perdew-Zunger
functional). In the literature,
we find use of both LDA\cite{son2006} and GGA\cite{huang2009} in the study of 
properties of graphene nanoribbons, and we expect that the choice of
exchange-correlation functional should not affect the main findings of our work much. 
We use an energy
cutoff for the plane-wave basis for wavefunctions (charge density) of 30 (240) Ry. 
Nanoribbons are simulated using a supercell geometry, with a vacuum layer of $>15$~\AA~between
any two periodic images of the GNR. A \textit{k}-point grid of 12 (24)x1x1 \textit{k} points
(periodic direction of the ribbon along $x$-axis) is used for sampling Brillouin zone
integrations for AGNR (ZGNR). 

Despite many-body effects in graphene being a subject of active research, 
most of the current experiments support validity of band structure point of view. 
Results of DFT calculations are also found to be in remarkable agreement 
with those of the Hubbard model\cite{palacios2007}, which takes into account 
onsite electron-electron interactions. While we believe that many-body effects 
will not drastically alter our results for structure and energetics of Stone-Wales
defects and associated warping, it would indeed be an interesting research problem 
to analyze electronic transport properties with many body corrections, 
which will not be addressed in this paper.

\section{Results and Discussion}
\subsection{SW Defects in Bulk Graphene}
\label{bg}
We first develop understanding of the stresses associated with SW defects in {\it bulk} graphene
(supercells shown in \fig{fig1}(a), (b) and (c)), and also benchmark our methodology through
comparison with earlier works. The rotated bond that creates the SW defect (marked in red) makes
an angle $\theta$ with the horizontal axis ($x$ axis).
Based on $\theta$, we classify the defects as parallel ($\theta$ = 0\textdegree) and
angled ($\theta\neq$ 0\textdegree) and denote by the symbol $SW_{\|}$ (see~\fig{fig1} (a))
and $SW_\angle$ (see~\fig{fig1} (b) and (c)). We express normal (shear) stress by $\sigma$ ($\tau$).
We use +ve (-ve) $\sigma$ to denote compressive (tensile) stress.
For bulk graphene, the stresses are in the units of eV/\AA$^{2}$, obtained by multiplying
the stress tensor components with the supercell length along $z$ direction.
We show the direction of planar stresses acting on a graphene supercell in \fig{fig1}(d).
For $SW_\|$ defect, $\sigma_x$, $\sigma_y$ and $\tau_{xy}$ are 0.40, -0.27 and 0 eV/\AA${^2}$
respectively. Under rotation $\theta$, stress tensor components transform as
\begin{equation}
\begin{gathered}
\sigma_x(\theta)=\frac{1}{2}(\sigma_x+\sigma_y)+\frac{1}{2}(\sigma_x-\sigma_y)cos~2\theta+\tau_{xy}sin~2\theta\\
\sigma_y(\theta)=\frac{1}{2}(\sigma_x+\sigma_y)-\frac{1}{2}(\sigma_x-\sigma_y)cos~2\theta-\tau_{xy}sin~2\theta\\
\tau_{xy}(\theta)=-\frac{1}{2}(\sigma_x-\sigma_y)sin~2\theta+\tau_{xy}cos~2\theta
\end{gathered}
\end{equation}
When $\theta$=60\textdegree, $\sigma_x$, $\sigma_y$ and $\tau_{xy}$ are -0.10,
0.23 and -0.29 eV/\AA${^2}$ respectively. $SW_\angle$ with $\theta$=120\textdegree~
is same as $\theta$=60\textdegree~in terms of stresses, barring the fact that
$\tau_{xy}$ has opposite sign. The energy cost for a single defect
formation in a 60 atom supercell is 5.4 eV (4.8 eV) for $SW_\|$ ($SW_\angle$), which is in
good agreement with Ref.~\onlinecite{lusk2008}. 
The energy difference is due to sizeable long range {\it anisotropic} interactions between the defects (periodic images),
and can be understood in the framework of Ref.~\onlinecite{ertekin2009}.

GNRs are obtained by cutting the bulk graphene sheet along certain direction $-$ along $x$ ($y$) to
create ZGNR (AGNR) (see \fig{fig1}) and the respective direction becomes the ribbon axis. Based on the analysis presented in the previous paragraph, we can readily predict the nature (sign) of stresses generated by SW defects in a GNR along the ribbon axis and in the transverse direction (i.e. along the ribbon width). However, due to finite thickness, a GNR has the freedom to relax stress along the width by deformation
(expansion/contraction depending on the sign of stress). We find that, in a properly relaxed SW
reconstructed GNR, except the normal stress acting along the ribbon axis, all the other stress tensor components are negligible. Thus, post structural relaxation, compressive (tensile) stress along the ribbon axis remains the only significant term in a $SW_\|$ reconstructed ZGNR (AGNR).
For $SW_\angle$ defect, the sign of induced stress along the ribbon axis is opposite to
that of $SW_\|$.

Based on bulk results, we can also predict the elastic energy cost of SW defect formation in
GNRs. Note that, normal stress created by $SW_\|$ in a particular direction is of higher magnitude
than that generated by $SW_\angle$ defect$-$ in $x$ direction, $\sigma_{\|}/\sigma_\angle=4$, and in $y$ direction,
$\sigma_{\|}/\sigma_\angle=1.2$. Elastic energy cost for defect formation is proportional to the stress. Hence, in a GNR $SW_\|$ defect is energetically more expensive than $SW_\angle$.
From the above discussion, it is evident that \textit{orientation of the defects with respect to the ribbon axis} plays a vital role in GNRs. We investigate this and its consequences in the rest of this paper.

\subsection{SW Defects in GNRs}
\label{gnr}
\subsubsection{Structural Stability}
\label{mp}
\begin{figure}
\subfigure[Pristine AGNR]{\epsfxsize=4.0truecm \epsfbox{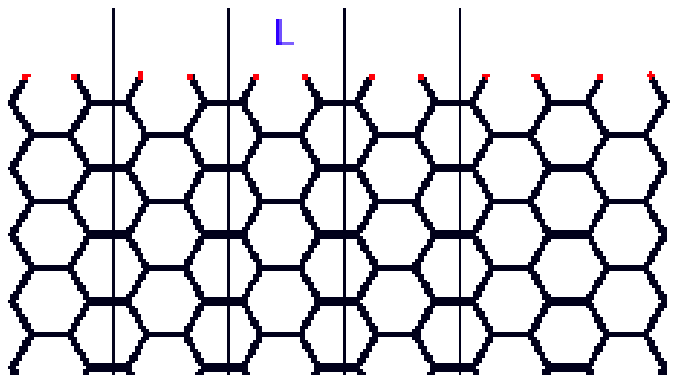}}
\subfigure[$A_{\bot 5757}^{2L}$]{\epsfxsize=4.0truecm \epsfbox{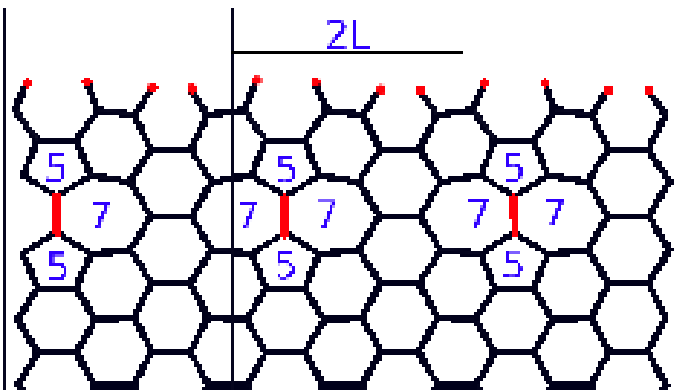}}
\subfigure[$A_{\bot 757}^{2L}$]{\epsfxsize=4.0truecm \epsfbox{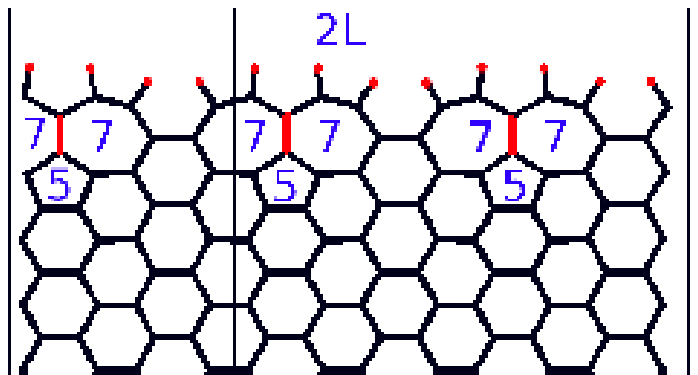}}
\subfigure[$A_{\angle 5757}^{2L}$]{\epsfxsize=4.0truecm \epsfbox{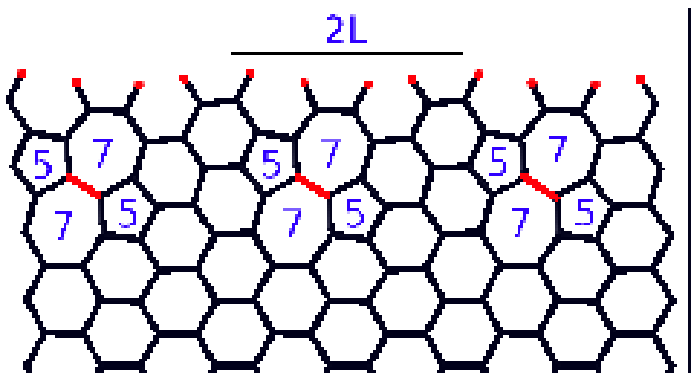}}
\subfigure[$A_{\angle 57}^{2L}$]{\epsfxsize=4.0truecm \epsfbox{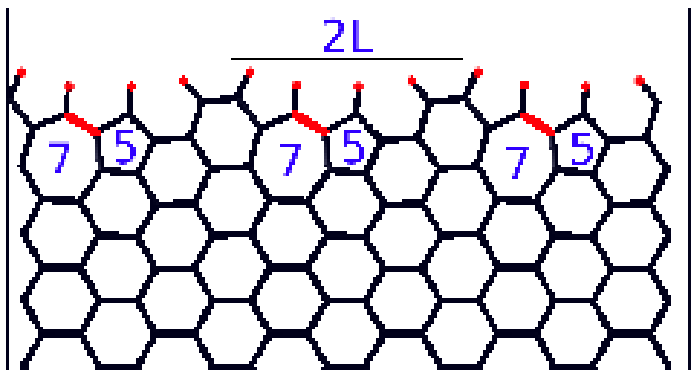}}
\subfigure[$A_{\angle 57}^{L}$]{\epsfxsize=4.0truecm \epsfbox{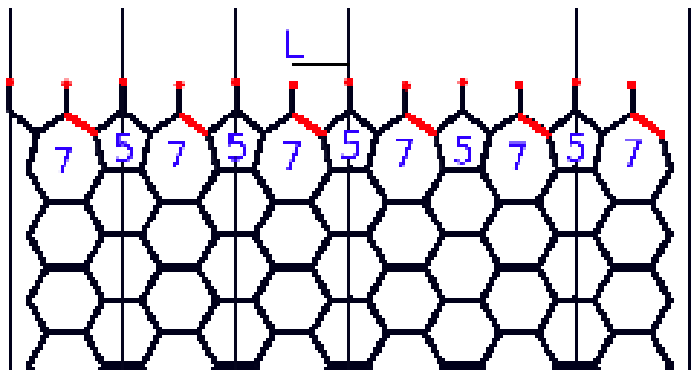}}
\subfigure[Pristine ZGNR]{\epsfxsize=4.0truecm \epsfbox{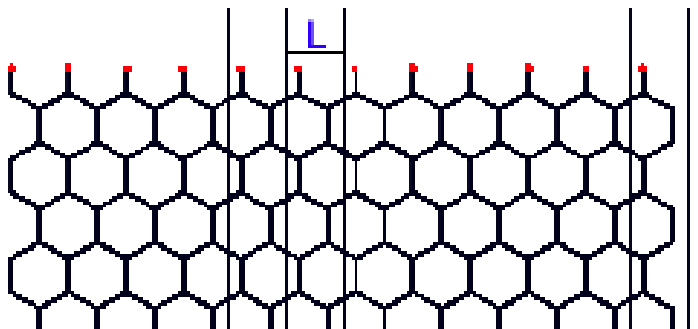}}
\subfigure[$Z_{\|575}^{3L}$]{\epsfxsize=4.0truecm \epsfbox{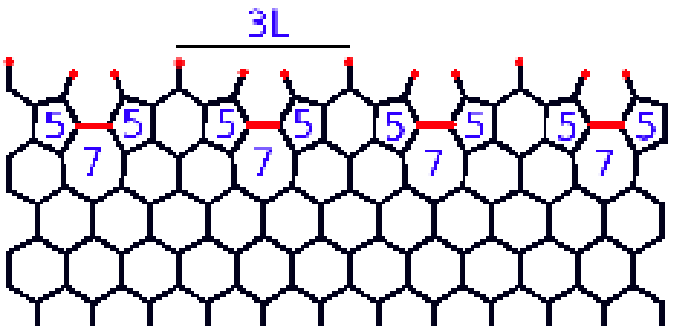}}
\subfigure[$Z_{\angle 5757}^{2L}$]{\epsfxsize=4.0truecm \epsfbox{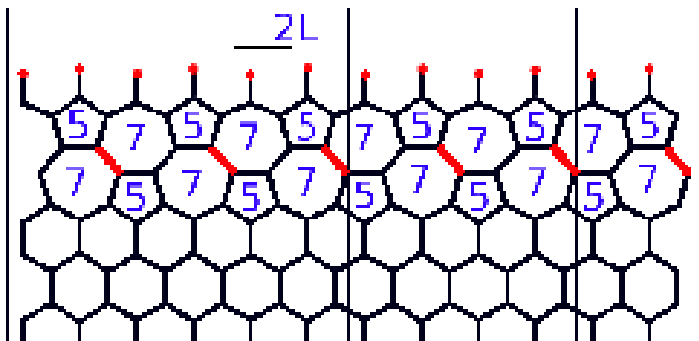}}
\subfigure[$Z_{\angle 57}^{2L}$]{\epsfxsize=4.0truecm \epsfbox{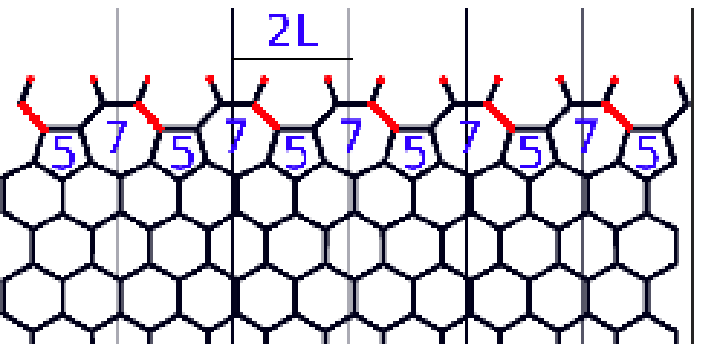}}
\subfigure[$Z_{\angle 5757}^{3L}$]{\epsfxsize=4.0truecm \epsfbox{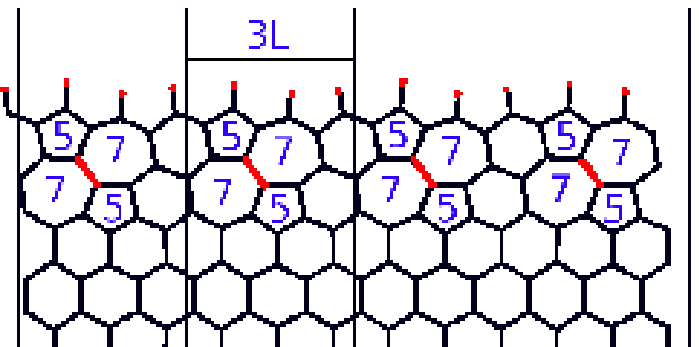}}
\subfigure[$Z_{\angle 57}^{3L}$]{\epsfxsize=4.0truecm \epsfbox{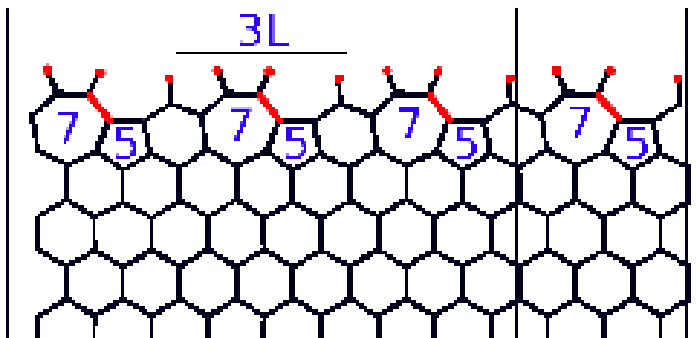}}
\caption{(color online) (a) Pristine AGNR and (b)$-$(f) SW reconstructed AGNRs. (g) Pristine ZGNR and (h)$-$(l) SW reconstructed ZGNRs. Red, appearing light gray in the gray-scale, dots and lines denote the hydrogen atoms and bonds rotated to create SW defects, respectively.}
\label{fig2}
\end{figure}
We first describe the nomenclature for different SW defects in GNRs.
The first letter, $A$ (armchair) or $Z$ (zigzag), denotes the kind of 
pristine GNR hosting a SW-defect, and the first subscript denotes the
orientation of SW defect, defined as the angle between the rotated bond and 
the ribbon axis. Three possible orientations are
$\|$ ($\theta=$ 0\textdegree), $\bot$ ($\theta=$ 90\textdegree) and
$\angle$ ($\theta\neq$ 0\textdegree ~or $\neq$ 90\textdegree). 
The $SW_\|$ defect in bulk graphene described earlier, falls 
into two categories in GNRs, $\bot$ and $\|$, to mark its
orientation with AGNR and ZGNR edge respectively. The series of 5's and 7's in the subscript
represent constituent rings of a single defect $-$ pentagons and heptagons. For example,
a $\bot$ defect, with a pair of pentagons and heptagons each in an AGNR is denoted as
$A_{\bot 5757}$ (see \fig{fig2}(b)). Such a defect is away from GNR edges and keeps the
armchair (or zigzag) edge shapes undisturbed (see~\fig{fig2}(b),(d),(i),(k)).
Lesser number of pentagons/heptagons ($575$ or $57$) in the
subscript implies defects overlapping with the edge, and reconstructed edge shapes typically
differ from that of pristine GNRs (see~\fig{fig2}(c),(e),(f),(h),(j),(l)). The superscript 
denotes the length of periodicity along the ribbon axis. $L=3d (\sqrt{3}d)$ for AGNR (ZGNR), where
$d$ is the C-C bond length. Finally, a subscript $w$ is used to differentiate warped ribbons 
from planar ones. All the GNRs reported here have H-terminated edges, shown by red dots in \fig{fig2}.
 
\begin{table}
\caption{Linear density of defects $\eta$ (number of SW defects per unit length), edge formation
energy $E_{edge}$, stress $\sigma$ along the ribbon axis and width W of various GNRs.}
\begin{center}
\begin{tabular}{llllllllllll}
\hline\hline
GNR&$\eta$&$E_{edge}$&\hspace{0.3cm}$\sigma$&W&GNR&$\eta$&$E_{edge}$&\hspace{0.3cm}$\sigma$&W\\
&\tiny{(/\AA)}&\tiny{(eV/\AA)}&\tiny{(eV/\AA)}&\tiny{(\AA)}& &\tiny{(/\AA)}&\tiny{(eV/\AA)}&\tiny{(eV/\AA
)}&\tiny{(\AA)} \\
\hline
AC                     &0    & 0.04 & 0    & 19.5 & Z                          &0   &0.10  & 0   & 15.5\\
$A_{\bot 5757}^{2L}$   &0.12 & 0.62 & -11  & 20.1 & $Z_{\parallel 575}^{3L}$   &0.14&0.86  & 19 & 15.0\\
$A_{\bot 757}^{2L}$    &0.12 & 0.51 & -11  & 20.0 & $Z_{\parallel 575w}^{3L}$  &0.14&0.70  & 7  & 14.8\\ 
$A_{\angle 5757}^{2L}$ &0.12 & 0.51 & 10   & 19.3 & $Z_{\parallel 557}^{2L}$   &0.21&1.66  & 29 & 14.8\\
$A_{\angle 57}^{L}$    &0.24 & 0.86 & 18   & 18.9 &$Z_{\parallel 557w}^{2L}$  &0.21&1.39  & 8  & 14.4\\
$A_{\angle 57w}^{L}$   &0.24 & 0.81 & 11   & 18.8 & $Z_{\angle 5757}^{3L}$     &0.14&0.49  & -5 & 15.9\\
$A_{\angle 57}^{2L}$   &0.12 & 0.40 & 7    & 19.3 & $Z_{\angle 5757}^{2L}$     &0.21&0.53  & -6 & 16.0\\
$A_{\angle 57w}^{2L}$  &0.12 & 0.36 & 3    & 19.0 & $Z_{\angle 57}^{3L}$       &0.14&0.34  & -4 & 16.0\\ 
                       &      &     &      &      & $Z_{\angle 57}^{2L}$       &0.21&0.37  & -5 & 16.1\\
\hline\hline
\end{tabular}
\end{center}
\label{t1}
\end{table}

We characterize GNRs with two properties: edge formation energy per unit length
and stress along the ribbon axis. The numerical values calculated using first
principles method are reported in Table~\ref{t1}. Edge formation energy per unit length is,
\begin{equation}
 E_{edge}=\frac{1}{2L}\left(E_{GNR}-N_CE_{bulk}-\frac{N_H}{2}E_{H_2}\right)
\end{equation}
where $E_{GNR}$, $E_{bulk}$ and $E_{H_2}$ are the total energies of the nanoribbon supercell,
one carbon atom in bulk graphene and of the isolated $H_2$ molecule respectively; $N_C$ ($N_H$) are
the number of carbon (hydrogen) atoms in the supercell. Stress reported here is 
$\sigma=(bc)\sigma_{x}$, where $\sigma_{x}$ is the component of stress tensor along
$x$ (ribbon axis) and $b$ and $c$ are the supercell sizes in $y$ and $z$ direction. Other
components of the stress tensor are negligible. $E_{edge}$ is found to be much higher for ZGNR
(0.10 eV/\AA) than compared to AGNR (0.04 eV/\AA). Our numbers are slightly overestimated with
respect to the reported values of 0.08 eV/\AA~(0.03 eV/\AA) for ZGNR (AGNR),\cite{wassmann2008} 
obtained using the PBE functional (a GGA functional) for exchange-correlation energy.

Edge defects, consisting of fewer number of pentagons and/or heptagons, require less formation energy. For example, consider $A_{\angle 5757}^{2L}$ and $A_{\angle57}^{2L}$, for which
$E_{edge}$ values are 0.51 and 0.40 eV/\AA~respectively. This is true for all SW reconstructed
GNRs if we compare the cases with defects of a particular orientation (see~Table~\ref{t1}).
For varied orientations, $\angle$ SW defects require less formation energy than compared to $\bot$ 
and $\|$ ones. For example, $E_{edge}$ of $A_{\angle 5757}^{2L}$ is lower by 0.11 eV/\AA~than that
of $A_{\bot 5757}^{2L}$ (consult~Table~\ref{t1} for more such instances). This observation is
consistent with the argument based on our analysis of SW defect in bulk graphene. Note that ribbons with higher linear defect density ($\eta$) have higher $E_{edge}$ and the above comparisons are
meaningful only for edge reconstructed GNRs of similar $\eta$.

Defect orientations with respect to the ribbon edges control the sign of stress induced.
Note that, the sign of stress along the ribbon axis reported in~Table~\ref{t1} matches with the predictions based on our analysis of SW defect in bulk graphene. The ribbon widths (W) vary in the range of 18.8 to 20.1 \AA~for AGNRs and 14.4 to 16.1~\AA~for ZGNRs. This is due to the stress relaxation via deformation in the direction perpendicular to the ribbon axis. For example, an unrelaxed $A_{\bot 5757}^{2L}$ experiences compressive stress along the width and relieves it by expansion in that direction; thus making it slightly wider than the pristine AGNR (consult~Table~\ref{t1}). Similarly, relaxation of the tensile stress along the width by contraction makes a SW reconstructed GNR narrower than pristine GNRs (see~Table~\ref{t1}).

\begin{figure}
\subfigure[$A_{\angle 57w}^L$]{\epsfxsize=8.5truecm \epsfbox{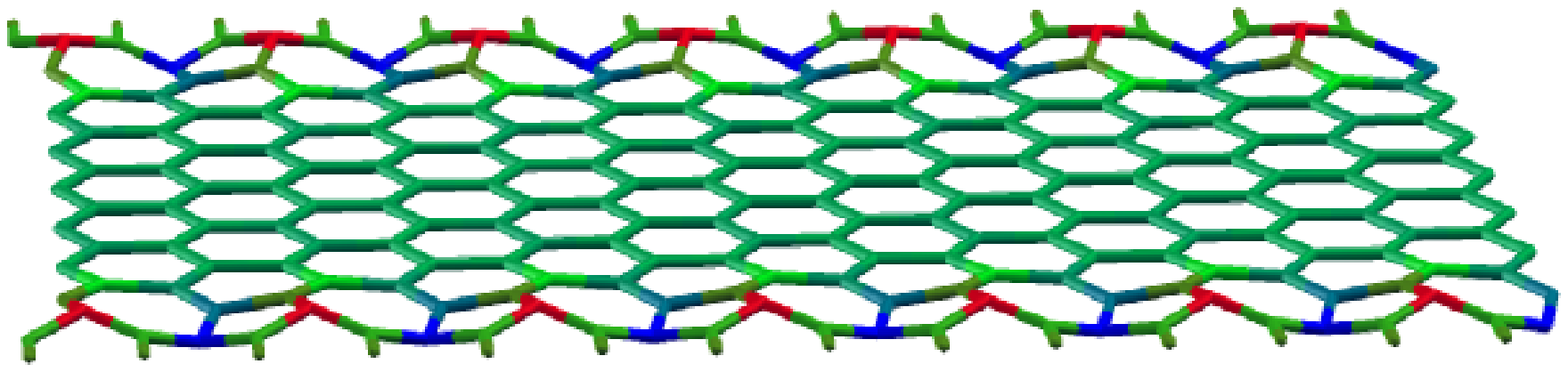}}
\subfigure[$A_{\angle 57w}^{2L}$]{\epsfxsize=8.5truecm \epsfbox{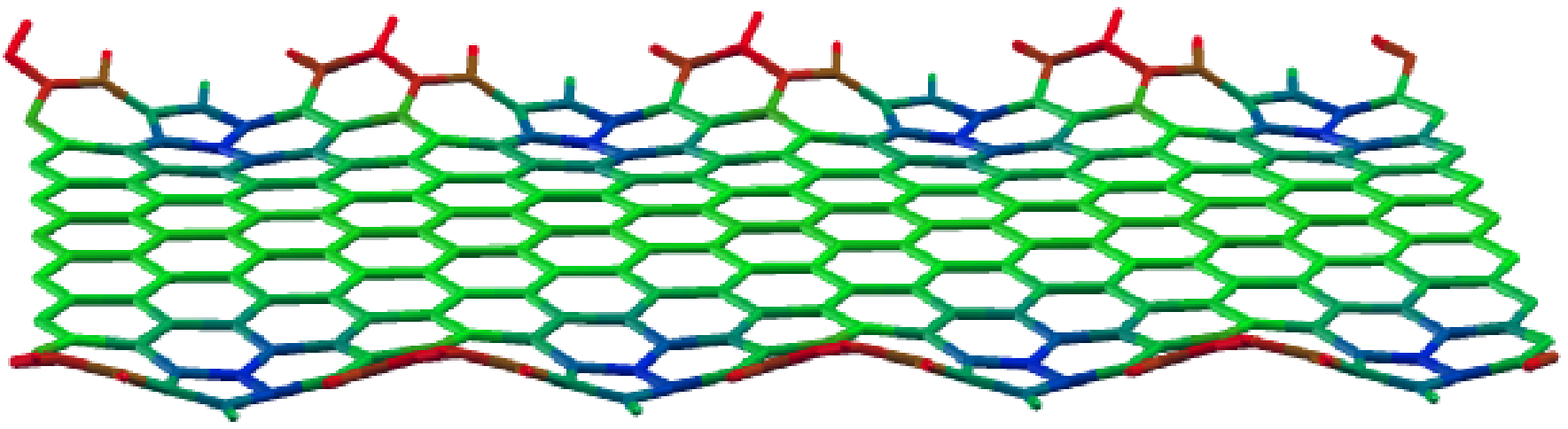}}
\subfigure[$Z_{\parallel 575w}^{3L}$]{\epsfxsize=8.5truecm \epsfbox{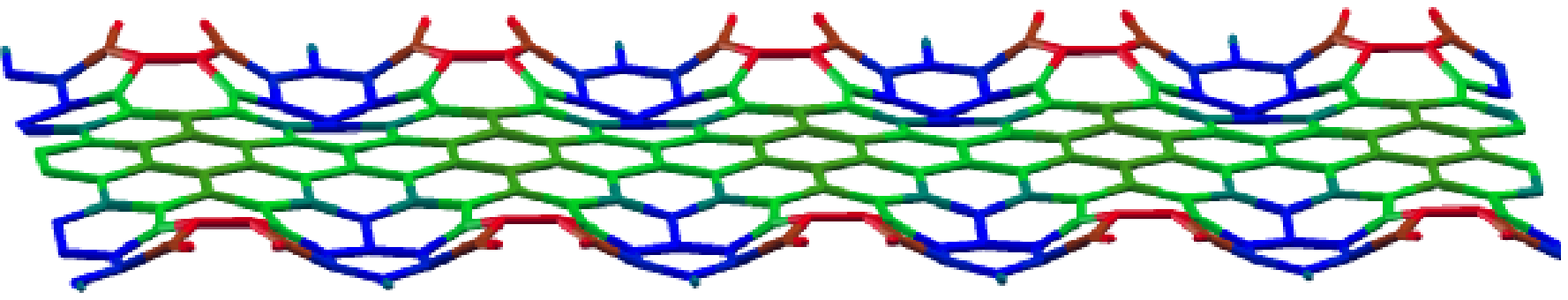}}
\caption{(color online) Warped structure of various different edge reconstructed nanoribbons.
Red, green and blue represents elevated, ``flat'' and depressed regions respectively.
In gray-scale, light and dark gray represents ``flat'' and warped regions respectively.
Atoms of a warped GNR are labeled as ``flat'' if $|z|<0.1$\AA, $z$ being
the height of the constituent carbon atoms. For a GNR of strictly flat geometry, $z=0$
for all the carbon atoms.}
\label{fig3}
\end{figure}

In contrast to the in-plane deformation to relax normal stress along the width,
the only way to partially release the normal stress along ribbon axis is by an out-of-plane 
deformation $-$ bending. It has been reported that pristine GNRs with dangling bonds relax the
compressive stress by spontaneous warping.\cite{shenoy2008} As shown in~\fig{fig3}, we also find
that SW reconstructed GNRs under compressive stress ($\angle$ ($\|$) in AGNR (ZGNR))
relax by local warping. For example, $\sigma$ ($E_{edge}$) is smaller 
by 4 (0.04) eV/\AA~ in $A_{\angle 57w}^{2L}$ than its planar form $A_{\angle 57}^{2L}$
(see~Table~\ref{t1} for more such instances). On the other hand, SW reconstructed GNRs under
tensile stress ($\bot$ ($\angle$) in AGNR (ZGNR)) favor planar structure. In this regard, our
results are in agreement with those of Ref.~\onlinecite{huang2009}, where authors have
considered only the SW defects generating tensile stress and found them to stabilize
the planar geometry. 

Our results for the edge reconstructions in $Z_{\angle 57}$ agree well with 
earlier work~\cite{wassmann2008}: our estimate of $E_{edge}$ (0.37 eV/\AA) 
for $Z_{\angle 57}^{2L}$ is slightly higher than the PBE (a GGA exchange-correlation
functional) estimate (0.33 eV/\AA) of Ref.~\onlinecite{wassmann2008}. However, 
we predict here qualitatively different types of reconstructions
in $A_{\angle 57}$ and $Z_{\|575}$ accompanied by warping of H-GNRs, 
which originate from different orientations of the SW-defects.
The gain in energy by H-saturation of dangling bonds at the edges of a GNR is so large
that even after the creation of a relatively costly SW-defect, it remains lower in energy
than the GNRs (no H-saturation) studied in Ref.~\onlinecite{huang2009} (a set of DFT
calculations, but the authors did not specify the exchange-correlation functional).
For example, SW-reconstructions at the edges (with dangling bonds) lead to a energy gain of 0.01 (0.18) eV/\AA~ for AGNR (ZGNR).\cite{huang2009} In contrast, we find SW reconstruction of H-GNRs
{\it costs} 0.32 (0.24) eV/\AA~ energy in AGNR (ZGNR) at least! However, because the 
edge energies of pristine H-GNR and GNRs with dangling bonds are of the order of 
0.1 eV/\AA~ and 1.0 eV/\AA~ respectively, $E_{edge}$ of the edge
reconstructed H-GNRs presented here (0.36 and 0.34 eV/\AA~ for AGNR and ZGNR)
is much smaller than that (0.99 and 0.97 eV/\AA~ for AGNR and ZGNR) reported in
Ref.~\onlinecite{huang2009}.

So far, we have presented theoretical analysis of Stone-Wales defects
in GNRs with a fixed width (19.5 and 15.5 \AA~ for pristine AGNR and ZGNR respectively).
Since interactions among the SW defects are long-ranged in nature, it will be interesting 
to verify how our findings depend on the width of a GNR.
$E_{edge}$ and $\sigma$ for edge reconstructed ribbons of widths 20.7 and 16.9 \AA, 
for pristine AGNR and ZGNR respectively, are found to be almost the same as values 
reported in Table~\ref{t1}, for GNRs with smaller width. Specifically, our estimates of
$E_{edge}$ and $\sigma$ for $A^L_{\angle 57w}$ type of edge reconstruction are 0.80 eV/\AA~ and 
11 eV/\AA~ respectively for wider ribbons. Changes in $E_{edge}$ with width of a GNR are similar
to those in a pristine GNR. Despite long-ranged interactions 
among the SW defects, such remarkable insensitivity to width can be understood in terms of defect concentration 
along the ribbon length vs. width. Since defects are located at the edges, distance between the two 
adjacent defects along the width is $15-20$ \AA~ (see the W column of Table~\ref{t1}). On the other hand, 
inter-defect distance along the ribbon length is about $4-8$ \AA~ (inverse of the number reported 
in the $\eta$ column of Table~\ref{t1}). Thus, defect-defect interactions along the length of the 
ribbon are dominant, explaining relatively weak dependence of edge properties on the width.  
We note that the GNRs used in experiments are typically wider than the ones we studied here,
and thus our results for edge reconstruction and related phenomena should hold good in such cases.

The ribbon periodicity ($L$, corresponding to minimum $E_{edge}$ and $\sigma$ in pristine GNRs)  was kept fixed in our analysis, as reported in Table~\ref{t1}. As shown in \fig{fig3}, for certain types of SW reconstructions, buckling relieves $\sigma$ partially and reduce $E_{edge}$. Nevertheless,
there is a small remanant stress, which could be relieved further by allowing the ribbons 
to relax along the periodic axis. For example, $E_{edge}$ and $\sigma$ decreases to 0.57 and 1.0 eV/\AA~ 
respectively, upon relaxation of the periodic length of ribbon $Z^{3L}_{\parallel 575w}$.
This results in an expansion of the ribbon by $4\%$ along its length. Warping still prevails, 
though with slightly smaller amplitude and longer wavelength. Thus our results do not change
qualitatively.
We note that relaxation of the periodicity of a GNR involves (a) an elastic energy cost associated with straining of the bulk (central part) of the ribbon and (b) a small energy gain associated with relief of
compressive stress at the edges. The former would dominate in wide ribbons typically used in
experiments, and our results obtained without such relaxation are more relevant to experimental
measurements.

Comparing the $E_{edge}$ values from~Table~\ref{t1}, we conclude that among the edge reconstructed
GNRs: \textit{warped AGNRs and flat ZGNRs are energetically more favorable} than flat AGNRs
and warped ZGNRs. In H-unsaturated GNRs, at an optimal concentration, SW reconstructions
lower the edge energy.\cite{huang2009} We also find that $E_{edge}$ values decrease on
reducing the linear defect density $\eta$ (by embedding SW defect in a longer supercell).
However, our study is limited to a region of high $\eta$ only. Whether there exists an
optimal $\eta$ in H-saturated GNRs also or not, at which reconstructed edge has lower
energy than pristine edge, needs to be investigated and is outside the scope of the present
paper.
\subsubsection{Electronic Properties}
\label{ep}
\begin{figure}
{\epsfxsize=8truecm \epsfbox{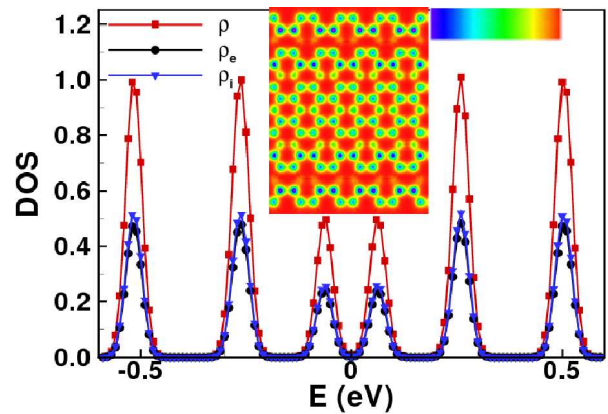}}
\caption{(color online) Density of states (DOS) and scanning tunneling microscope (STM) image of
pristine AGNR visualized with XCRYSDEN~\cite{kokalj}. $E=0$ denotes the Fermi energy. 
We have applied a thermal broadening equivalent to room temperature to plot the DOS in this figure and
throughout the rest of the paper. Consult text for the definition of $\rho$, $\rho_e$ and $\rho_i$.
The horizontal bar represents the color scale used in the STM image. The  image has been simulated for
a sample bias of -0.3 eV and reflects the spatial distribution of the local density of states LDOS
below Fermi energy.}
\label{fig4}
\end{figure}
Electronic properties of GNRs are sensitive to the geometry of the edges at the boundary. For example,
pristine AGNRs are semiconducting - the bandgap arises due to quantum confinement and depends
on the width of the ribbon.\cite{son2006} Pristine ZGNRs also exhibit gapped energy spectrum, although
of entirely different origin. Gap arises due to a localized edge potential generated by
edge magnetization.\cite{son2006} It is well known that presence of defects or disorders at
the edges can change the electronic, magnetic and transport properties of GNRs to various
extent.\cite{kumazaki,schubert} In 2D graphene, topological defects such as dislocations or
SW defects give rise to electronic states localized at the defect sites.\cite{ana2008}
Presence of any such defect induced states near the edges of the SW reconstructed GNRs can have interesting consequences on their electronic and transport properties. 
 
In this section, we analyze electronic properties using the density of states (DOS) and simulated 
scanning tunneling microscope (STM) images of pristine and edge reconstructed GNRs.
We decompose total DOS ($\rho$) into the sum of projected
DOS of atoms located at the interior of the ribbon ($\rho_i$) and atoms near the edges ($\rho_e$).
$\rho_e$ is the sum of projected DOS of first two layers of atoms from both the edges.
Since the defects are located at the edges, this technique clearly uncovers the difference between
electronic band structures of a pristine and edge reconstructed GNR.
Note (\fig{fig3}) that,
this is the region which undergoes warping (remains flat) if the ribbon edges are under compressive
(tensile) stress. $\rho_i$ includes the projected DOS of rest of the atoms (located in the region 
of the nanoribbon that always remains flat). Depending on the sample bias, STM images help identify the 
spatial distribution of local DOS (LDOS) below (-ve bias) and above (+ve bias) the Fermi level ($E_F=0$). These images
should be useful in experimental characterization of GNRs, as well as understanding consequences of such defects
and warping to electronic transport in GNRs.

The DOS for a 19.5~\AA~wide pristine AGNR of bandgap 0.1 eV (\fig{fig4})
shows that both $\rho_i$ and $\rho_e$ contributes to the $\rho$ with equal weight, and
symmetric about $E_F$. This symmetry is known as particle-hole symmetry. The
inset of~\fig{fig4} shows that for a sample bias of -0.3 eV, the occupied LDOS
is spread over the entire ribbon. We do not present the corresponding image for a
positive sample bias, which is very similar due to the underlying particle-hole symmetry.

\begin{figure}
{\epsfxsize=8truecm \epsfbox{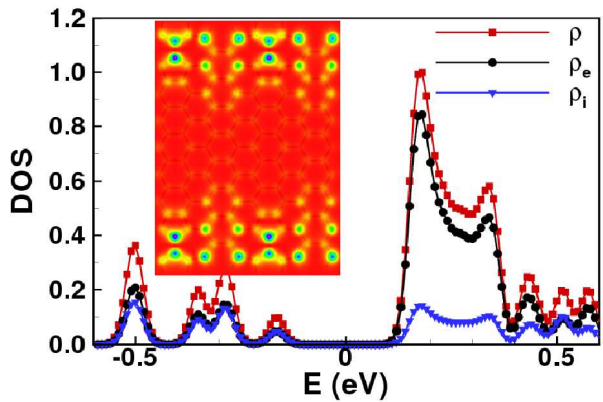}}
\caption{(color online) DOS and STM image of $A_{\bot 757}^{2L}$, simulated with a sample bias of +0.2 eV. 
The STM image shows the spatial distribution of LDOS above the Fermi energy.}
\label{fig5}
\end{figure}

Edge reconstruction by $\bot$ SW defect ($7-5-7$ defect) breaks the particle-hole symmetry and a sharp peak of DOS
appears above $E_F$ (see~\fig{fig5}), which has primary contribution from the edge atoms ($\rho_e>\rho_i$).
The STM image (simulated with a sample bias of +0.2 eV) reveals that
the unoccupied LDOS is localized at the defect sites, located very close to the ribbon edges.
The STM image for a -ve sample bias (not shown here) illustrates that occupied LDOS is spatially distributed over the
entire ribbon, similar to the pristine one.

\begin{figure}
\subfigure[$A_{\angle 57}^{L}$]{\epsfxsize=8truecm \epsfbox{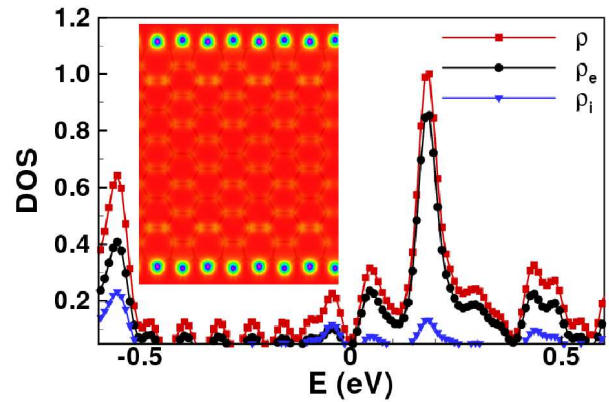}}
\subfigure[$A_{\angle 57w}^{L}$]{\epsfxsize=8truecm \epsfbox{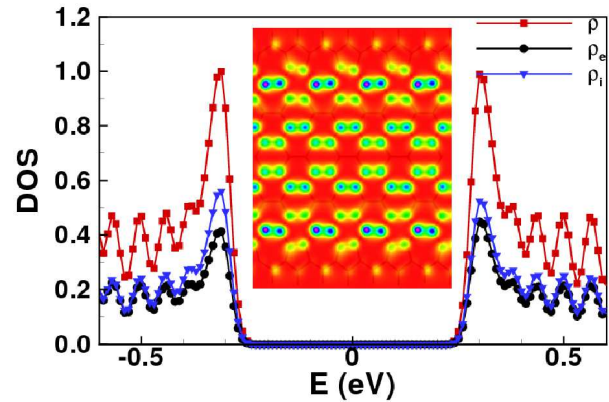}}
\caption{(color online) DOS and STM image of (a) $A_{\angle 57}^{L}$ and (b) $A_{\angle 57w}^{L}$.
The STM images have been simulated with sample bias of +0.2 and +0.4 eV for (a) and (b) respectively.
They show the spatial distribution of LDOS above the Fermi energy.}
\label{fig6}
\end{figure}

Edge reconstruction by $\angle$ SW defects ($5-7$ defect) have similar consequences on electronic band structure of 
AGNRs. A sharp peak in DOS, primarily coming from $\rho_e$, appears above the Fermi energy 
(see~\fig{fig6}(a)). Simulated STM image (at sample bias +0.2 eV) confirms that LDOS above the Fermi level
are localized at the edges. For -ve sample bias (not presented here), occupied LDOS is spatially
distributed over edge, as well as interior atoms. However, for this type of edge reconstructions,
planar structure is not the stable one and the ribbon undergoes warping near the edges (see~\fig{fig3}).
As shown in~\fig{fig6}(b), the DOS peak of edge-localized (or rather defect localized) states above the Fermi energy
vanishes in the warped GNR and LDOS above $E_F$ is spatially distributed throughout the ribbon.
This is also true for LDOS below the Fermi level also(STM image not shown here).
This reveals an electronic origin of the defect induced stress and its anisotropy in the localized $p$
like defect state. Delocalization of this state relieves the stress and favors warping.

All the results presented here are for a pristine AGNR of narrow bandgap
(0.1 eV), which undergoes edge reconstructions by various SW defects. We have investigated
edge reconstructions of a wide bandgap AGNR also. We find qualitative similarity, such as unoccupied
LDOS localized at the edges, among the edge reconstructed AGNRs of various widths and bandgaps.
However, the magnitude of bandgap depends both on ribbon width and the type of SW reconstruction
present at the edge (i.e. edge shape) and varies over a wide range of values (0.1 eV to 1 eV).
The unoccupied states localized near the edges can have interesting applications in molecule
detection. These states are going to act as electron acceptors and can detect some suitable
electron donating molecules.

\begin{figure}
{\epsfxsize=8truecm \epsfbox{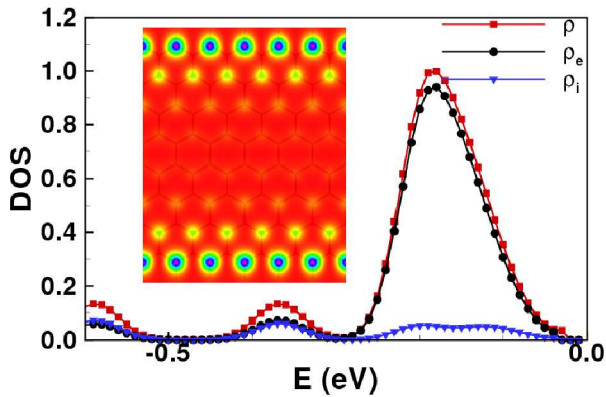}}
\caption{(color online) DOS up to $E_F$(=0.0) and STM image of pristine ZGNR, simulated with a sample bias
of -0.2 eV. The image illustrates the spatial distribution of LDOS below the Fermi energy.}
\label{fig7}
\end{figure}

We have employed a local spin density approximation (LSDA)
in our calculations to explore the possibilies of magnetic ordering in GNRs.
We initialize our calculation with atoms at the edges of GNRs to  
have non-zero spin polarization (of same (opposite) sign at the two edges in
FM (AFM) configurations). While the magnitude of spin at edge atoms change
in the course to self-consistency, their relative signs remain the same, if 
the corresponding magnetic ordering is stable. As mentioned earlier, pristine ZGNRs
have gapped antiferromagnetic ground sate.\cite{son2006,young2006} We illustrate the DOS
and simulated STM image of a pristine ZGNR with a width of 15.5~\AA~in~\fig{fig7}.
The bandgap is 0.3 eV and we show DOS upto the Fermi level. Note that up and down spin electrons
have similar energy spectrum and are not shown separately. The STM image has been
simulated for a sample bias of -0.2 eV and reveals that LDOS below the Fermi energy is localized
at the zigzag edges. The edges are spin polarized - ferromagnetically coupled along a particular
edge but antiferromagnetic between two opposite edges (not shown here).

\begin{figure}
{\epsfxsize=8truecm \epsfbox{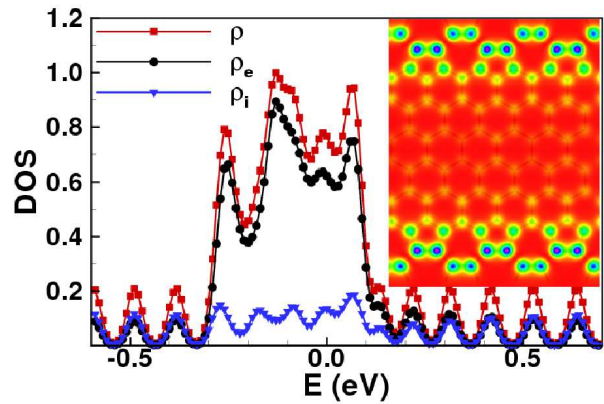}}
\caption{(color online) DOS and STM image of $Z_{\angle 57}^{2L}$, simulated with a sample bias of
-0.2 eV. The image shows the spatial distribution of occupied LDOS.}
\label{fig8}
\end{figure}

We find that edge reconstructions by SW defects \textit{destroy magnetism}. At high defect density,
this leads to a~\textit{nonmagnetic metallic} ground state and at lower defect
density magnetism survives with a weaker magnitude. In this paper, our investigation is
restricted to the regime of high defect density (where all or most of the zigzag edges have been
reconstructed by SW defects - see~\fig{fig2}(h)-(l)) and we do not discuss the issue of magnetism any further.
The states at $E_F$ arise primarily from the edges for all the zigzag GNRs.

The DOS and simulated STM image (bias voltage -0.2 eV) of a $\angle$ SW ($5-7$ defect) edge reconstructed ZGNR (see \fig{fig8})
reveal a nonzero DOS at $E_F$(=0) and that the ground state is of a nonmagnetic metal. The STM image shows the formation
of nearly isolated dimers along the edge. Reconstruction of ZGNR edge by $\parallel$ SW defect
($5-7-5$ defect) does not alter the electronic properties qualitatively. Such ZGNRs are also
nonmagnetic metallic with planar as well as warped geometries (not shown here). However, these are very high energy
edges and are unlikely to be preferred over $5-7$ SW reconstructions in ZGNRs.
The $5-7$ defects can act as interface of hybrid graphene
and hybrid GNRs, having both armchair and zigzag like features. Such materials have
remarkable electronic and magnetic properties.~\cite{botello2009}

\section{Conclusion}
\label{con}
In conclusion, the sign of stress induced by a SW-defect in a GNR depends on 
the orientation of the SW-defect with respect to the ribbon edge, and the relaxation of the 
structure to relieve this stress drives its stability. Local warping or wrinkles 
arise in the GNR when the stress is compressive, while the structure remains planar otherwise.
The specific consequences to AGNR and ZGNR can be understood from the anisotropy of the 
stress induced by a SW defect embedded in bulk graphene. Using the analogy between a SW-defect and a dislocation, it should be possible to capture the interaction between a SW defect in the interior of a GNR and its edge within a continuum framework that includes images of SW-defects in the edges. As the images of SW-defects are also SW-defects, their interactions can be readily captured within the continuum framework of Ref.~\onlinecite{ertekin2009}. Our work shows how warping of GNRs can be nucleated at the SW-defects localized at the edges and be responsible for flake-like 
shapes of graphene samples seen commonly in experiments. Such warping results in delocalization of electrons in the defect states. In ZGNRs, magnetic ordering weakens due to the presence of SW defects at the edges
and the ground state is driven towards that of a nonmagnetic metal.

\section{Acknowledgment}
SB thanks Vijay B Shenoy for valuable discussions, suggestions and comments. UVW acknowledges 
support from an IBM Faculty award.

\bibliographystyle{apsrev}
\bibliography{ref.bib}

\end{document}